\documentclass[onecolumn,notitlepage,showpacs]{revtex4-1}

\usepackage{amsmath}
\usepackage{amsfonts}
\usepackage{amssymb}
\usepackage{bm}
\usepackage{bbm}
\usepackage{nicefrac}
\usepackage{cancel}
\usepackage{xcolor}
\usepackage{maybemath}
\usepackage{pifont}
\usepackage{afterpage}
\usepackage{subfloat}
\usepackage{siunitx}

\usepackage{dcolumn}
\usepackage{pifont} 

\usepackage[%
colorlinks=true,
urlcolor=blue,
linkcolor=blue,
citecolor=blue
]{hyperref}

\def\ii{{\mathrm{i}}}

\def\vdw{van der Waals}

\def\calU{\mathcal{U}}

\def\calH{{\mathcal H}}
\def\calL{{\mathcal L}}
\def\calF{{\mathcal F}}
\def\calO{{\mathcal O}}
\def\calV{{\mathcal V}}
\def\calW{{\mathcal W}}

\def\HFS{{\mathrm{HFS}}}
\newcommand{\fs}{{\mathrm{FS}}}

\def\LS{{\mathrm{LS}}}
\def\vdW{{\mathrm{vdW}}}

\def\DHF{{\calH}}
\def\DDD{{\calV}}
\def\WWW{{\calW}}
\def\FS4{{\calF}}
\def\FSF{{\calF}}

\newcommand{\lrgrule}{\rule[-4mm]{0mm}{11mm}}
\newcommand{\stdrule}{\rule[-2mm]{0mm}{6mm}}

\makeatletter

\newcommand{\Rmnum}[1]{\expandafter\@slowromancap\romannumeral #1@}

\newcolumntype{.}{D{x}{}{-1}}

\definecolor{duelferred}{rgb}{0.7, 0.2, 0.1}

\begin{document}

\title{\texorpdfstring{$\bm{6P}$--$\bm{1S}$}{6P--1S} Interaction}

\title{Long--Range Interactions for Hydrogen: 
\texorpdfstring{$\bm{6P}$--$\bm{1S}$ and $\bm{6P}$--$\bm{2S}$}{6P--1S and 6P--2S} 
Systems}

\author{U. D. Jentschura and C. M. Adhikari}

\affiliation{Department of Physics,
Missouri University of Science and Technology,
Rolla, Missouri 65409, USA}

\begin{abstract}
The collisional shift of a transition constitutes
an important systematic effect in high-precision 
spectroscopy. Accurate values for \vdw{} interaction coefficients
are required in order to evaluate the distance-dependent 
frequency shift. We here consider the interaction of 
excited hydrogen $6P$ atoms with metastable atoms
(in the $2S$ state), in order to explore the 
influence of quasi-degenerate $2P$, and $6S$ states
on the dipole-dipole interaction.
The motivation for the calculation is given 
by planned high-precision measurements of 
the transition. Due to the presence of quasi-degenerate
levels, one can use the non-retarded approximation 
for the interaction terms over wide distance ranges.
\end{abstract}


\maketitle


%
%
\section{Introduction}
\label{sec1}

The long-range interaction of identical atoms,
one of which is in an excited state,
constitutes an interesting physical problem~\cite{Ch1972,DeYo1973}.
This is mainly due to energetic degeneracies 
connected with the ``exchange'' of the states 
among the two atoms. For $nS$--$1S$ interactions (atomic 
hydrogen), this problem has recently been investigated in 
Ref.~\cite{AdEtAl2017vdWiii}. It was found that the 
interesting oscillatory $1/R^2$
long-range tails~\cite{SaKa2015,Be2015,MiRa2015,DoGuLa2015,Do2016,JeAdDe2017prl,%
JeDe2017vdw0}
are numerically suppressed and become dominant only for 
excessively large interatomic distances, 
in a region where the absolute magnitude of the 
interaction terms is numerically insignificant.
Indeed, the Casimir--Polder regime of a $1/R^7$ interaction
is never reached for systems with at least one atom 
in an excited state~\cite{JeAdDe2017prl,JeDe2017vdw0}.

A completely different situation is encountered 
when both atoms are in excited states, or, when the 
excited state is accessible from the ground state 
via an allowed electric-dipole transition~\cite{JeEtAl2017vdWii}.
In this case, one encounters nonvanishing first-order 
\vdw{} interaction matrix elements instead of second-order effects.
We recall that the \vdw{} Hamiltonian for the 
interaction of atoms $A$ and $B$ reads as 
[in SI MKSA (meter-kilogram-second-Ampere) units],
\begin{equation}
\label{HVDW}
H_{\rm vdW} = 
\frac{1}{4\pi\,\epsilon_0}\frac{ \beta_{ij} \, d_{Ai} \, d_{Bj} }{R^3} \,,
\qquad
\beta_{ik} = \delta_{ik} - 3 \frac{R_i \, R_k}{R^2} \,,
\end{equation}
where $\vec R = \vec R_A - \vec R_B$, 
$R = | \vec R|$   $\hat{R}=\vec{R}/R$.
Furthermore, $\vec d_{A}=e\, \vec r_{A}$ is the electric dipole moment 
operator for atom $A$  and  $\vec d_{B}$ is the same for atom $B$
($\vec r_i$ with $i=A,B$ is the electron coordinate
relative to the atomic nucleus).
States of the form, e.g., $| 6P \rangle_A \, | 1S \rangle_B$
are energetically degenerate with respect to 
states of the form $| 1S \rangle_A \, | 6P \rangle_B$,
and are coupled by the \vdw{} Hamiltonian.
For the $2S$--$2P$ system, this problem 
has been analyzed in Ref.~\cite{JoEtAl2002},
on the basis of nonrelativistic Schr\"{o}dinger theory.
However, in order to evaluate the distance-dependent 
frequency shift of hyperfine-resolved 
transitions, one has to invoke a more sophisticated 
analysis, which has recently been performed in 
Ref.~\cite{JeEtAl2017vdWii}.
A complete rediagonalization of the 
total Hamiltonian, comprising Lamb shift,
fine-structure, and hyperfine effects, becomes necessary.

In order to analyze the problem, one has to 
define a quantization axis, which we choose as the 
line of separation of the two atoms. 
This brings the \vdw{} Hamiltonian into the form
\begin{equation}
\label{HvdW}
H_{\rm vdW} = \frac{e^2}{4 \pi \epsilon_0} \,
\frac{x_A \, x_B + y_A \, y_B - 2 \, z_A \, z_B}{R^3}  \,.
\end{equation}
We should add that the interaction remains 
non-retarded over very wide distance ranges,
commensurate with the fine-structure and 
Lamb shift transition wavelengths in the quasi-degenerate system.

Here, we engage in the endeavor of evaluating 
long-range interactions for the 
hydrogen $6P$--$1S$ and $6P$--$2S$ systems. In the latter case, 
we find it necessary to include, in our basis of states,
all $2S$, $2P$, $6S$ as well as $6P_{1/2}$ and $6P_{3/2}$ 
hyperfine-resolved atomic levels.
The calculations are motivated, in part,
by the prospect of a future high-precision measurement
of the $2S$--$6P$ transition in atomic
hydrogen~\cite{UdHaPriv2017},
to supplement ongoing efforts for a resolution
of the proton charge radius puzzle~\cite{PoEtAl2010,AnEtAl2013,PoEtAl2016}
(see the recent work~\cite{BeEtAl2017} for a 
discussion of systematic effects in $4P$--$2S$ hydrogen
systems, which are closely related to the systems under investigation here).

%
%
\section{General Formalism}
\label{sec2}

The total Hamiltonian is 
\begin{equation}
\label{H}
H = H_{\LS,A} + H_{\LS,B} +
H_{\HFS,A} + H_{\HFS,B} + 
H_{\fs,A} + H_{\fs,B} + 
H_{\vdW} \,,
\end{equation}
where $H_{\rm LS}$ stands for the Lamb shift,
$H_{\rm HFS}$ describes hyperfine effects, and 
$H_{\fs}$ describes fine-structure splittings.
These Hamiltonians have to be added for atoms $A$ and $B$,
\begin{subequations}
\begin{align}
\label{HHFS}
H_{{\rm HFS}, i} =& \; \frac{\mu_0}{4\pi}\mu_B\mu_N\,
g_s \, g_p
\left[\frac{8\pi}{3}\vec{S}_i\cdot\vec{I}_i\,\delta^{(3)}\left(\vec{r}_i\right)
+\frac{3 (\vec{S}_i\cdot\vec{r}_i) \,
(\vec{I}_i\cdot\vec{r}_i) -
\vec{S}_i\cdot\vec{I}_i \; \left|\vec{r}_i\right|^{\,2}}{\left|\vec{r}_i\right|^5} 
+ \frac{\vec{L}_i\cdot\vec{I}_i}{\left|\vec{r}_i\right|^3} \right] \,,
\\
\label{HLS}
H_{{\rm LS},i} =& \; \frac{4}{3}
\frac{\hbar^3 \, \alpha^2 }{m_e^2 \, c} \,
\left(\frac{\hbar}{m_e c}\right)^3\ln\left(\alpha^{-2}\right)
\delta^{(3)}\left(\vec{r}_i\right) \,,
\\
\label{HFS}
H_{{\rm FS},i} =& \; - \frac{\vec p_i^{\,4}}{8 m_e^3 \, c^2} 
+ \frac{\pi \, \hbar^3 \, \alpha}{2 \, m_e^2 \, c} \, \delta^{(3)}(\vec r_i) 
+ \frac{\hbar^3 \, \alpha}{4 m_e^2 \, c \, r^3} \vec \sigma \cdot \vec L \,.
\end{align}
\end{subequations}
Here, $i = A,B$ denotes either atom, while
$\alpha$ is the fine-structure constant
and $m_e$ is the electron mass.
The momentum operators for the atomic electrons 
are denoted as $\vec p_i$,
while $\vec{L}_i$ is the orbital angular momentum operator.
The (dimensionless) spin operator for electron $i$
is $\vec{S}_i$, while $\vec{I}_i$ is the spin operator for proton $i$.
The CODATA (Committee on Data for Science and Technology)  
values (see Ref.~\cite{MoNeTa2016})
for the electronic and protonic $g$ factors are
$g_s\simeq \num{2.002319}$ and $g_p\simeq \num{5.585695}$.
The Bohr magneton is 
$\mu_B\simeq \SI{9.274010e-24}{\ampere\metre^2}$,
while $\mu_N\simeq \SI{5.050784e-27}{\ampere\metre^2}$ 
is the nuclear magneton. 
The expression for $H_{\rm LS}$ in Eq.~(\ref{HLS})
follows the Welton approximation~\cite{ItZu1980},
which is sufficient for our purposes of 
calculating long-range interaction coefficients.

For the $6P$--$1S$ and $6P$--$2S$ systems, we 
define  the zero of the energy 
as the sum of the 
Dirac energies of the $1S$ and $6P_{1/2}$ states,
and to the sum of the $2S$ and $6P_{1/2}$ states,
respectively.
The zero point of the energy excludes 
both Lamb shift as well as hyperfine effects;
in the following, we add the Lamb shift energy to 
the $S$ states but leave the $P$ states untouched by
Lamb shift effects. So, 
our definition of the zero point of the energy 
corresponds to the hyperfine centroid of the 
$|(6P_{1/2})_A (1S)_B \rangle$ states,
and to the hyperfine centroid of the
$|(6P_{1/2})_A (2P_{1/2})_B \rangle$ system,
respectively. The fine-structure energy is added 
for the $6P_{3/2}$ states.

The matrix elements of the \vdw{} Hamiltonian have to
be calculated in a hyperfine-resolved basis.
Let us take the $6P_{1/2}$--$1S$ interaction 
as an example, and exclude the $6P_{3/2}$ states
for the time being. The unperturbed states carry the 
quantum numbers
\begin{subequations}
\begin{align}
1S_{1/2}(F=0): & \; n=1, \, \ell = 0, \, J =\frac12, \, F=0 \,, \\
1S_{1/2}(F=1): & \; n=1, \, \ell = 0, \, J =\frac12, \, F=1 \,, \\
6P_{1/2}(F=0): & \; n=6, \, \ell = 1, \, J =\frac12, \, F=0 \,, \\
6P_{1/2}(F=1): & \; n=6, \, \ell = 1, \, J =\frac12, \, F=1 \,.
\end{align}
\end{subequations}
Here, the quantum numbers have their usual 
meaning, i.e., $n$ is the principal 
quantum number, while $\ell$, $J$, and $F$,
respectively, are the 
electronic orbital angular momentum, the
total (orbital$+$spin) electronic angular momentum,
and the total (electronic$+$protonic) atomic angular momentum.
The multiplicity of the hyperfine-resolved state 
is $g_F = 2F+1$.

After adding the electron orbital and 
spin angular momenta, and the nuclear (proton) spin $| \pm \rangle_p$,
the four $1S$ states within the hyperfine manifold
are given by
\begin{subequations}
\label{1S:states}
\begin{align}
\left\vert n=1, \ell = 0, J=\frac12, F=0, F_{z}=0 \right\rangle 
=& \; -\frac{
\left\vert +\right\rangle_p \,
\left\vert -\right\rangle_e -
\left\vert -\right\rangle_p \,
\left\vert +\right\rangle_e}{\sqrt{2}} \;
\left\vert 1, 0,0\right\rangle_e \,, \\
\left\vert n=1,\ell = 0,J=\frac12, F=1,F_{z}=0 \right\rangle 
=& \; \frac{
\left\vert + \right\rangle_p \,
\left\vert -\right\rangle_e +
\left\vert -\right\rangle_p \,
\left\vert +\right\rangle_e}{\sqrt{2}}\left\vert 1,0,0\right\rangle_e 
\,, \\
\left\vert n=1,\ell = 0,J=\frac12, F=1,F_{z}=\pm1 \right\rangle 
=& \; \left\vert \pm\right\rangle_p \,
\left\vert \pm\right\rangle_e \,
\left\vert 1,0,0\right\rangle_e \,,
\end{align}
\end{subequations}
while the hyperfine singlet $6P_{1/2}$ ($F=0$) state is given by 
\begin{align}
\left\vert n=6,\ell=1,J=\frac12, F=0,F_{z}=0 \right\rangle 
=& \; \frac{1}{\sqrt{3}} 
\left\vert +\right\rangle_p 
\left\vert +\right\rangle_e 
\left\vert 6,1,-1\right\rangle_e  
- \frac{1}{\sqrt{6}} 
\left\vert +\right\rangle_p 
\left\vert -\right\rangle_e 
\left\vert 6,1,0\right\rangle_e
\nonumber\\[0.1133ex]
& \; + \frac{1}{\sqrt{3}} 
\left\vert -\right\rangle_p 
\left\vert -\right\rangle_e
\left\vert 6,1,1\right\rangle_e 
- \frac{1}{\sqrt{6}} 
\left\vert -\right\rangle_p 
\left\vert +\right\rangle_e
\left\vert 6,1,0\right\rangle_e \,.
\end{align}
The hyperfine triplet states in the $6P$ manifold read as follows,
\begin{align}
\left\vert n=6, \ell=1,J=\frac12,F=1,F_{z}=0 \right\rangle
=& \; -\frac{1}{\sqrt{3}}
\left\vert +\right\rangle_p 
\left\vert +\right\rangle_e
\left\vert 6,1,-1\right\rangle_e 
+\frac{1}{\sqrt{6}}
\left\vert +\right\rangle_p 
\left\vert -\right\rangle_e
\left\vert 6,1,0\right\rangle_e 
\nonumber\\[0.1133ex]
& \; + \frac{1}{\sqrt{3}}
\left\vert -\right\rangle_p 
\left\vert -\right\rangle_e 
\left\vert 6,1,1\right\rangle_e -
\frac{1}{\sqrt{6}} 
\left\vert -\right\rangle_p
\left\vert +\right\rangle_e
\left\vert 6,1,0\right\rangle_e \,,
\end{align}
and
\begin{equation}
\label{6Pstates}
\left\vert n=6,\ell=1,J=\frac12, F=1,F_{z}=\pm1 \right\rangle 
= \mp\frac{1}{\sqrt{3}} 
\left\vert \pm\right\rangle_p 
\left[ 
\left\vert \pm\right\rangle_e 
\left\vert 6,1,0\right\rangle_e\right.
 -\sqrt{2}
\left.\left\vert \mp\right\rangle_e \,
\left\vert 6,1,\pm1\right\rangle_e \right].
\end{equation}
Here and in the following, we use the notation $| n,\ell, J, F, F_z \rangle$ for the 
unperturbed states with the specified quantum numbers.

\begin{table*}[t!]
\begin{center}
\begin{minipage}{0.7\linewidth}
\caption{\label{table1} Multiplicities in the $6P_{1/2}$--$6P_{3/2}$--$1S$ system.
One might wonder why $F_z = \pm 3$ is possible for $F=2$. The
answer is that $F=2$ here refers to the total angular momentum
(electron orbital plus electron spin plus nuclear spin)
of one of the atoms, while $F_z = \pm 3$ refers to the
angular momentum projection of the sum of the
total angular momenta of both electrons.}
\begin{tabular}{ccccc}
\hline
\hline
\stdrule
                & $F_z=0$   &   $F_z=\pm 1$ &    $F_z=\pm 2$ &   $F_z=\pm 3$ \\
\hline
\hline
\stdrule
$(J=\frac32,F=2)$   &    8       &      8    &       6    &      2     \\
\stdrule
$(J=\frac32,F=1)$   &    8       &      6    &       2    &      0     \\
\hline
\stdrule
$(J=\frac32)$       &    16      &     14    &       8    &     2    \\
\hline
\stdrule
$(J=\frac12,F=1)$   &     8      &      6    &       2    &      0     \\
\stdrule
$(J=\frac12,F=0)$   &     4      &      2    &       0    &      0     \\
\hline
\stdrule
$(J=1/2)$       &    12      &      8    &       2    &      0     \\
\hline
\hline
\stdrule
$(J=\frac12)+(J=\frac32)$  & 28      &     22    &      10    &      2     \\
\hline
\hline
\end{tabular}
\end{minipage}
\end{center}
\end{table*}

%
%
\section{\texorpdfstring{$\bm{6P}$--$\bm{1S}$}{6P--1S} Interaction}
\label{sec3}

%
%
\subsection{Orientation}
\label{sec3A}

We need to diagonalize the Hamiltonian
given in Eq.~(\ref{H}),
in a quasi-degenerate basis, composed of two 
atoms, the first being in a $6P$ state,
the second being in a substate of the $1S$ hyperfine manifold.
Retardation does not need to be considered.
For the manifolds composed of the $(6P)_A (1S)_B$ and 
$(1S)_A (6P)_B$ states, with all hyperfine levels
resolved, we obtain the following total multiplicities 
when all $6P_{1/2}$ and $6P_{3/2}$ states
are added into the basis (see also Table~\ref{table1}):
$g(F_z = \pm 3) = 2$, $g(F_z = \pm 2) = 10$,
$g(F_z = \pm 1) = 22$, $g(F_z = 0) = 28$.
The multiplicities are the sums of the multiplicities in the 
$6P_{3/2}$--$1S$ and the $6P_{1/2}$--$1S$ system.

We now turn to the computation of the matrix elements of the total 
Hamiltonian~(\ref{H}) in the space spanned 
by the two-atom states which are product states
built from any two states of the types
given in Eqs.~(\ref{1S:states}) (for the $1S$ states)
and~(\ref{6Pstates}) (for the $6P_{1/2}$ states),
as well as the $6P_{3/2}$ states.
Matrix elements of the 
\vdw{} interaction Hamiltonian are determined 
with the help of a computer symbolic program~\cite{Wo1999}.
We define the parameters
%
%
\begin{subequations}
\begin{align}
\label{parameters}
\calH \equiv & \;
\frac{\alpha^4}{18} \, g_p \, \frac{m_e }{m_p}\,m_e \,c^2 
= h \, 59.21498 \, {\rm MHz} \,,
& \calL_2 & \equiv h \, 1057.845(9) \, {\rm MHz} \,, \\
\calL_6 \equiv & \;
h \times \frac{1}{27}\, \times1057.845(9)\, {\rm MHz} \,,
& \calF & \equiv \frac{\alpha^4 \, m_e  \, c^2}{864} 
= h \, 405.529\, {\rm MHz} \,, 
\\
\calV \equiv & \; 3\, \frac{e^2}{4 \pi \epsilon_0} \,
\frac{a_0^2}{R^3} = \frac{ 3 \, E_h }{ \rho^3 } \,,
& \calW & \equiv 
\frac{2^{15} \times 3^5 \times 5^7}{7^{17}} \, 
\frac{e^2}{4 \pi \epsilon_0} \, \frac{a_0^2}{R^3} = 
\frac{2^{15} \times 3^5 \times 5^7}{7^{17}} \, 
\frac{ E_h }{ \rho^3 } \,,
\end{align}
\end{subequations}
where $R = a_0 \rho$, and $a_0$ is the Bohr radius.
$E_h=\alpha^2 m_e c^2$ is the Hartree energy.
Our scale $\calH$ is equal 
to one-third of the hyperfine splitting of 
the $2S$ state~\cite{KoEtAl2009},
while $\calL_2$ is the $2S$-$2P_{1/2}$ Lamb shift~\cite{LuPi1981},
and $\calL_6$ approximates the $6S$-$6P_{1/2}$ Lamb shift.
The natural scale for the constants $\calH$ and $\calL$ is 
an energy of order $\alpha^3 \, E_h$. Hence, we write
\begin{equation}
\calH = \alpha^3 \, E_h \, C_{\calH} \,,
\qquad
\calL_n =  \alpha^3 \, E_h \, C_{\calL,n} \,,\qquad
\calF = \alpha^3 \, E_h \, C_{\calF} \,,
\end{equation}
where we set $C_{\calH}= 0.0231596$,  
$C_{\calF} = 0.158606$ and  $C_{\calL,2}= 27 \, C_{\calL,6} = 0.413734$.
Typical second-order energy shifts encountered in
our calculations may be expressed as
\begin{equation}
\frac{\calV^2}{T_1 \calH + T_2 \calL_n +T_3\calF} 
=\frac{9}{T_1 \, C_{\calH} + T_2 \,C_{\calL,n}+T_3C_{\calF}} \, \,
\frac{E_h}{\alpha^3 \,\rho^6} \,,
\end{equation}
where $T_1$, $T_2$  and $T_3$ are numerical coefficients
of order unity.

Particular attention should be devoted 
to mixing terms of the Hamiltonian,
in the space of the $6P_{1/2}$ and $6P_{3/2}$ states,
within the $F=1$ manifolds,
with the mixing matrix element being given by
$\langle 6P_{3/2}^{F=1} (F_z) | H_{\rm HFS} |
6P_{1/2}^{F=1} (F_z)  \rangle = X$
(see Ref.~\cite{Pa1996mu} for an outline of the 
calculation).

In order to illustrate the mixing term,
we temporarily restrict the discussion here to one atom only,
say, atom $A$, omitting the subscript on 
$H_{\rm HFS} \equiv H_{{\rm HFS},A}$.
For the two states, with the same magnetic projection $F_z = \mu$,
we consider the basis of states
\begin{equation}
| a \rangle = | 6P_{1/2}^{F=1} (F_z = \mu) \rangle 
 = | 6, 1, \frac12, 1, \mu \rangle \,, 
\qquad
| b \rangle = | 6P_{3/2}^{F=1} (F_z = \mu \rangle 
= | 6, 1, \frac32, 1, \mu  \rangle \,, 
\qquad
\mu = -1,0,1 \,.
\end{equation}
The matrix of the Hamiltonian $H_{\rm HFS} + H_{\rm FS}$
in the basis $\{ | a \rangle, | b \rangle \}$ is evaluated as
\begin{equation}
\label{HFSFSmix}
H_{\rm HFS+FS} = 
\left( \begin{array}{cc}
D & X \\
X & -D+\calF \\
\end{array} \right) \,,
\qquad
D = g_p \frac{\alpha^4 m_e^2c^2}{1944 \, m_p} \,,
\qquad
X = -g_p \frac{\alpha^4 m_e^2  c^2}{3888 \, \sqrt{2} \, m_p} \,.
\end{equation}
Here, $g_p$ is the proton $g$ factor,
while $D$ is a diagonal matrix element, and $X$ is the off-diagonal
element given above.
The eigenvalues of $H^{F_z=1}_{\rm HFS+FS}$ are given by
\begin{equation}
\mathcal{E}_{+} = -D +\calF +\frac{X^2}{\calF-2D} +\calO(X^4)\,,
\qquad
\mathcal{E}_{-} = D-\frac{X^2}{\calF-2D} +\calO(X^4).
\end{equation}
The second-order shift in the eigenvalues, 
$\Delta = X^2/(\calF-2D)$, is numerically 
equal to $ \num{1.412129e-13} E_h$, 
where $E_h= \alpha^2 m_ec^2$ is the Hartree energy.
For simplicity, we thus define the parameter
\begin{equation} \label{DefineDelta}
\Delta = \num{1.412129e-14} \,,
\qquad
\frac{\Delta \cdot E_h}{h} = \SI{92.9137}{\hertz} \,.
\end{equation}
The hyperfine splitting energy between $6P_{1/2}(F=1)$ and 
$6P_{1/2}(F=0)$ states thus amounts to $\mathcal{H}/27$, 
while between $6P_{3/2}(F=2)$ and $6P_{3/2}(F=1)$ states,
it is $2 \mathcal{H}/135$.
The $1S$-state hyperfine splitting is $24 \mathcal{H}$.
For the product state of atoms $A$ and $B$,
we shall use the notation
$| (n_A, \ell_A, J_A, F_A, F_{z,A})_A \, 
(n_B, \ell_B, J_B, F_B, F_{z,B})_B \, \rangle$,
which summarizes the quantum numbers of both atoms.

%
%
\subsection{States with \texorpdfstring{$\bm{F_z = 3}$}{Fz = 3}}
\label{sec3B}

The atomic states can be classified according to the 
quantum number $F_z$; the $z$ component of the 
total angular momentum 
commutes \cite{JeEtAl2017vdWii} with the total Hamiltonian
given in Eq.~(\ref{H}).
Within the $6P_{1/2}$--$6P_{3/2}$--${1S}_{1/2}$ system,
the states in the manifold $F_z = 3$ are given as follows,
\begin{equation}
| \phi_{1} \rangle = | (1,0,\frac12, 1,1)_A \, (6,1, \frac32, 2,2)_B \rangle \,,
\quad
| \phi_{2} \rangle = | (6,1,\frac32,2,2)_A \, (1,0,\frac12,1,1)_B \rangle \,.
\end{equation}
Here, we have ordered the basis vectors in ascending order of
the quantum numbers, starting from the last member in 
the list. The Hamiltonian matrix evaluates to 
\begin{equation}
\label{matFz2}
H_{F_z = 3} = 
\left(
\begin{array}{cc}
\dfrac{1081}{180} \,\DHF{} + \calF & 
\dfrac{2^{15} \times 3^5 \times 5^7}{7^{17}} \DDD{} \\[4ex]
\dfrac{2^{15} \times 3^5 \times 5^7}{7^{17}} \DDD{} & 
\dfrac{1081}{180} \,\DHF{} + \calF \\
\end{array}
\right) \,.
\end{equation}
We have subtracted the sum of the Dirac energies of the 
$1S$ and $6P_{1/2}$ hyperfine centroids, and the $1S$ Lamb shift 
is absorbed in the definition of the $1S$ hyperfine centroid
energy, as outlined in Sec.~\ref{sec2}.
The energy eigenvalues and eigenvectors 
corresponding to $H_{F_z = 3}$ are given as follows,
\begin{equation}
\label{eigenenergy}
E_{\pm} =
\frac{1081}{180} \DHF{} + \calF\mp 
\dfrac{2^{15} \times 3^5 \times 5^7}{7^{17}}  \DDD{} \,,
\qquad
| u_\pm \rangle = \frac{1}{\sqrt{2}} \, 
( | \phi_{1} \rangle \pm | \phi_{2} \rangle ) \,.
\end{equation}
The average of the first-order shifts (linear in $\DDD{}$) vanishes;
there are no second-order shifts (quadratic in $\DDD{}$).

%
%
\subsection{Manifold \texorpdfstring{$\bm{F_z = 2}$}{Fz = 2}}
\label{sec3C}

We order the $10$ states in this manifold in 
order of ascending quantum numbers, 
%
\begin{subequations}
\label{statesFz2}
\begin{align}
| \psi_{1} \rangle = & \; | (1,0,\frac12,0,0)_A \, (6,1,\frac32,2,2)_B \rangle
\,,
& | \psi_{2} \rangle & = | (1,0,\frac12,1,0)_A \, (6,1,\frac32,2,2)_B \rangle
\,, \\[0.1133ex]
| \psi_{3} \rangle = & \; | (1,0,\frac12,1,1)_A \, (6,1,\frac12,1,1)_B \rangle
\,, 
& | \psi_{4} \rangle & = | (1,0,\frac12,1,1)_A \, (6,1,\frac32,1,1)_B \rangle
\,, \\[0.1133ex]
| \psi_{5} \rangle = & \; | (1,0,\frac12,1,1)_A \, (6,1,\frac32,2,1)_B \rangle
\,,  
& | \psi_{6} \rangle & = | (6,1,\frac12,1,1)_A \, (1,0,\frac12,1,1)_B \rangle
\,, \\[0.1133ex]
| \psi_{7} \rangle = & \; | (6,1,\frac32,1,1)_A \, (1,0,\frac12,1,1)_B \rangle
\,,  
& | \psi_{8} \rangle & = | (6,1,\frac32,2,1)_A \, (1,0,\frac12,1,1)_B \rangle
\,, \\[0.1133ex]
\label{average}
| \psi_{9} \rangle = & \; | (6,1,\frac32,2,2)_A \, (1,0,\frac12,0,0)_B \rangle
\,, 
& | \psi_{10} \rangle & = | (6,1,\frac32,2,2)_A \, (1,0,\frac12,1,0)_B \rangle \,.
\end{align}
\end{subequations}
States $| \psi_{3} \rangle$ and $| \psi_{6} \rangle$ are $6P_{1/2}$ states,
the rest are $6P_{3/2}$ states (see also the multiplicities indicated in 
Table~\ref{table1}). Among the $6P_{3/2}$ states, 
$| \psi_{4} \rangle$ and $| \psi_{7} \rangle$ have $F = 1$, the rest have 
$F = 2$. The Hamiltonian matrix 
$\calH = H_{F_z = 2}$ is $10$-dimensional,
\begin{equation}
\begin{small}
\label{matFz1}
\calH = 
\left(
\begin{array}{cccccccccc}
 \FSF-\frac{3239 \DHF}{180} & 0 & 0 & 0 & 0 
& -\sqrt{3} \WWW & \sqrt{6} \WWW & 0 & 0 & 0 \\
0 & \frac{1081 \DHF}{180}+\FSF & 0 & 0 & 0 & 
\sqrt{3} \WWW & \sqrt{\frac{3}{2}} \WWW & \frac{3 \WWW}{\sqrt{2}} & 0 & 0 \\
0 & 0 & \frac{649 \DHF}{108} & -\frac{\DHF}{216 \sqrt{2}} & 0 & 
-2 \WWW & -\sqrt{2} \WWW & \sqrt{6} \WWW & -\sqrt{3} \WWW & \sqrt{3} \WWW \\
0 & 0 & -\frac{\DHF}{216 \sqrt{2}} & \frac{647 \DHF}{108}+\FSF & 
0 & -\sqrt{2} \WWW & -\WWW & \sqrt{3} \WWW & \sqrt{6} \WWW & \sqrt{\frac{3}{2}} \WWW \\
0 & 0 & 0 & 0 & \frac{1081 \DHF}{180}+\FSF & \sqrt{6} \WWW & \sqrt{3} \WWW & 
-3 \WWW & 0 & \frac{3 \WWW}{\sqrt{2}} \\
-\sqrt{3} \WWW & \sqrt{3} \WWW & -2 \WWW & -\sqrt{2} \WWW & \sqrt{6} \WWW & 
\frac{649 \DHF}{108} & -\frac{\DHF}{216 \sqrt{2}} & 0 & 0 & 0 \\
\sqrt{6} \WWW & \sqrt{\frac{3}{2}} \WWW & -\sqrt{2} \WWW & 
-\WWW & \sqrt{3} \WWW & -\frac{\DHF}{216 \sqrt{2}} &
\frac{647 \DHF}{108}+\FSF & 0 & 0 & 0 \\
0 & \frac{3 \WWW}{\sqrt{2}} & \sqrt{6} \WWW & \sqrt{3} \WWW & 
-3 \WWW & 0 & 0 & \frac{1081 \DHF}{180}+\FSF & 0 & 0 \\
0 & 0 & -\sqrt{3} \WWW & \sqrt{6} \WWW & 0 & 0 & 0 & 0 & 
\FSF-\frac{3239 \DHF}{180} & 0 \\
0 & 0 & \sqrt{3} \WWW & \sqrt{\frac{3}{2}} \WWW & 
\frac{3 \WWW}{\sqrt{2}} & 0 & 0 & 0 & 0 & \frac{1081 \DHF}{180}+\FSF \\
\end{array}
\right).
\end{small}
\end{equation}
An adjacency graph~\cite{JeEtAl2017vdWii,AdDeJe2017aphb}
shows that the matrix is irreducible,
which in particular implies that there are no hidden 
symmetries in the  Hamiltonian matrix $H_{F_z = 2}$  which would otherwise 
lead to a further decomposition into irreducible submatrices. 

Of particular interest are second-order \vdw{} shifts,
within the ($F_z=2$) manifold. 
We have the quantum number $J$, $F$, and $F_z$ at our disposal.
A weighted average over the possible values of $F$, 
namely, $F=1$ and $F=2$, keeping $J = 1/2, 3/2$ and $F_z = +2$
fixed, leads to the result
\begin{subequations}
\begin{align}
\left< E(6P_{1/2}, F_z=2) \right>_F =& \;
\left( -\Delta - \dfrac{\num{1.13849e3}}{\rho^6} \right) E_h \,, \\
\left< E(6P_{3/2}, F_z = 2) \right>_F =& \;
\left( \frac14 \Delta + \dfrac{\num{2.84623e2}}{\rho^6} \right) \, E_h \,.
\end{align}
\end{subequations}
The weighted average vanishes,
\begin{equation}
2 \left< E(6P_{1/2}, F_z = 2) \right>_F +
8 \left< E(6P_{3/2}, F_z = 2) \right>_F = 0 \,.
\end{equation}
Observe that the $\Delta$ term, which is the HFS--FS 
(hyperfine-structure–fine–structure)  mixing 
term, only occurs for the $F=1$ states, and vanishes
for the $F=2$ states (see also the entries in Table~\ref{table2}).
Furthermore, observe that there are no
$6P_{1/2}$ states with $F=0$ in the manifold $F_z = 2$, because
of angular momentum selection rules (we have $F_z = 2$ 
and hence $F \geq 1$ for all states in the manifold). 

%
%
%
\subsection{States with \texorpdfstring{$\bm{F_z = 1,0,-1,-2,-3}$}{Fz = 1,0,-1,-2,-3}}
\label{sec3D}

We would like to list some of the $22$ states in the $F_z = 1$ manifold in
order of ascending quantum numbers,
$| \Psi_{1} \rangle =\;| (1,0,\frac12,0,0)_A \, (6,1,\frac12,1,1)_B \rangle$,
$| \Psi_{2} \rangle =\;| (1,0,\frac12,0,0)_A \, (6,1,\frac32,1,1)_B \rangle$,
$| \Psi_{3} \rangle =\;| (1,0,\frac12,0,0)_A \, (6,1,\frac32,2,1)_B \rangle$,
$| \Psi_{4} \rangle =\;| (1,0,\frac12,1,-1)_A \, (6,1,\frac32,2,2)_B \rangle$,
and so on, up to 
$| \Psi_{21} \rangle =\;| (6,1,\frac32,2,1)_A \, (1,0,\frac12,1,0)_B \rangle$
and 
$| \Psi_{22} \rangle =\;| (6,1,\frac32,2,2)_A \, 
\allowbreak (1,0,\frac12,1,-1)_B \rangle$.
We refer to Table~\ref{table2}
for the averaged second-order \vdw{} shifts in the $F_z = 0$, 
$F_z = +1$, $F_z = +2$,  and $F_z = +3$ manifolds.
The Hamiltonian matrix for  $F_z = -3$ manifold is 
identical to that of $F_z = +3$. The  $F_z = -2$  manifold has  identical 
diagonal entries to that of $F_z = +2$ while 
some off-diagonal entries are different. The same is true of the
$F_z=\pm1$ manifolds. Yet, we have checked that 
the distance-dependent Born--Oppenheimer energy curves for $F_z = \pm2$ 
and $F_z=\pm 1$ are alike.

\begin{table}[t!]
\begin{center}
\begin{minipage}{0.95\linewidth}
\begin{center}
\caption{\label{table2} Second-order \vdw{} shifts for $6P_J$ hydrogen atoms 
interacting with ground-state hydrogen 
atoms. Entries marked with a long hyphen
(--) indicate unphysical combinations of $F$ and $F_z$ values.
We denote the scaled interatomic distance by $\rho = R/a_0$ 
and give all energy shifts in atomic units, i.e., 
in units of the Hartree energy $E_h = \alpha^2 m_e c^2$. 
Recall that $F_z=F_{z,A}+F_{z,B}$.
The results still involve an averaging over the 
quantum numbers of the spectator atom;
e.g., the result for $J=3/2$, $F = 2$, and $F_z = 2$
involves and averaging over the second-order 
shifts of the states $| \Psi_9 \rangle$ and $| \Psi_{10} \rangle$
given in Eq.~\eqref{average}.
We define $\Delta$ in Eq.~(\ref{DefineDelta}).}
%
\begin{tabular}{c@{\hspace{0.5cm}}c@{\hspace{0.5cm}}c@{\hspace{0.5cm}}c@{\hspace{0.5cm}}c}
\hline
\hline
\stdrule
                & $F_z=0$   &   $F_z= \pm 1$ &    $F_z=\pm 2$ &   $F_z= \pm 3$ \\
\hline
\hline
\lrgrule
$(J=3/2,F=2)$   & $\dfrac{\num{1.07795e5}}{\rho^6}$ 
                & $\dfrac{\num{8.75949e4}}{\rho^6}$ 
                & $\dfrac{\num{8.06860e4}}{\rho^6}$ &  0 \\
\lrgrule
$(J=3/2,F=1)$   &  $\Delta-\dfrac{\num{1.06590e5}}{\rho^6}$ 
                &  $\Delta-\dfrac{\num{1.15780e5}}{\rho^6}$ 
                &  $\Delta-\dfrac{\num{2.40919e5}}{\rho^6}$ &  --- \\
\lrgrule
$(J=1/2,F=1)$   & $-\Delta + \dfrac{\num{9.99477e3}}{\rho^6}$
                & $-\Delta + \dfrac{\num{3.52047e4}}{\rho^6}$
                & $-\Delta - \dfrac{\num{1.13849e3}}{\rho^6}$ & --- \\
\lrgrule
$(J=1/2,F=0)$   & $-\dfrac{\num{2.23999e4}}{\rho^6}$ 
                & $-\dfrac{\num{1.08654e5}}{\rho^6}$ 
                & ---  &  --- \\
\hline
\hline
\end{tabular}
\end{center}
\end{minipage}
\end{center}
\end{table}

%
%
%
\subsection{Second--Order Energy Shifts}
\label{sec3E}

As a function of $J$ and $F$, 
within the $6P$--$1S$ system, 
a global averaging over all the states with
different individual magnetic quantum numbers
leads to the results
\begin{subequations}
\begin{align}
\left< E(6P_{1/2}, F=0 \right>_{F_z} =& \;
-\dfrac{\num{6.55265e4}}{\rho^6} \, E_h \,, \\
\left< E(6P_{1/2}, F=1 \right>_{F_z} =& \; 
\left( -\Delta + \dfrac{\num{2.07442e4}}{\rho^6} \right) E_h \,, \\
\left< E(6P_{3/2}, F=1 \right>_{F_z} =& \; 
\left( \Delta - \dfrac{\num{1.33573e5}}{\rho^6} \right) E_h \,, \\
\left< E(6P_{3/2}, F=2) \right>_{F_z}
=& \; \dfrac{\num{8.08028e4}}{\rho^6} \, E_h \,.
\end{align}
\end{subequations}
Comparing to Table~\ref{table2}, this average 
would correspond to an average over the entries in the 
different rows, for given~$F$
(for the multiplicities, see Table~\ref{table1}).

One can also average over the possible orientations of $F$,
namely, $F = J \pm \frac12$, for given $J$ and $F_z$.
This amounts to an averaging over the first two entries in the
columns, and the third and fourth entry in every column,
of Table~\ref{table2}.
The results are
\begin{subequations}
\begin{align}
\left< E(6P_{1/2}, F_z=0) \right>_F = & \;
\left( -\frac23  \Delta - \frac{\num{8.03454e2}}{\rho^6} \right) E_h \,, \\
\left< E(6P_{1/2}, F_z=\pm 1 \right>_F = & \;
\left( -\frac34 \Delta - \frac{\num{7.59733e2}}{\rho^6} \right) E_h , \\
\left< E(6P_{1/2}, F_z=\pm 2) \right>_F = & \;
\left( -\Delta - \frac{\num{1.13849e3}}{\rho^6} \right) E_h \,, 
\end{align}
\end{subequations}
and
\begin{subequations}
\begin{align}
\left< E(6P_{3/2}, F_z=0) \right>_F =& \;
\left( \frac12 \, \Delta + \dfrac{\num{6.02591e2}}{\rho^6} \right) E_h \,, \\
\left< E(6P_{3/2}, F_z=\pm 1 \right>_F =& 
\left( \frac37 \, \Delta + \dfrac{\num{4.34133e2}}{\rho^6} \right) E_h \,, \\
\left< E(6P_{3/2}, F_z=\pm 2) \right>_F =& 
\left( \frac14 \, \Delta + \dfrac{\num{2.84623e2}}{\rho^6} \right) E_h \,, \\
\left< E(6P_{3/2}, F_z=\pm 3 \right>_F =& \; 0 \; \,.
\end{align}
\end{subequations}
As a function of $J$, averaging over $F$ and $F_z$ leads to the
results
\begin{subequations}
\begin{align}
\label{6P1S_6P12}
\left< E(6P_{1/2} \right>_{F, F_z} =& \; 
\left( -\frac34 \, \Delta - \dfrac{\num{8.23474e2}}{\rho^6} \right) E_h \,, \\
\label{6P1S_6P32}
\left< E(6P_{3/2} \right>_{F, F_z} =& \; 
\left( \frac38 \, \Delta + \dfrac{\num{4.11737e2}}{\rho^6} \right) E_h \,.
\end{align}
\end{subequations}
Without hyperfine resolution, there are four $J = \frac32$ states
and two $J = \frac12$ states. Hence, the fine-structure 
average of the latter two results vanishes.

\begin{table*}
\begin{center}
\begin{minipage}{0.8\linewidth}
\begin{center}
\caption{\label{table3} 
Multiplicities in the $6P_{1/2}$--$6P_{3/2}$--($6S$;$2P_{1/2}$)--$2S$--$1S$ system.
The entries in the first seven rows refer to the 
$6P_{1/2}$--$6P_{3/2}$--$2S$ system, and are the same as those for 
the $6P_{1/2}$--$6P_{3/2}$--$1S$ system given in Table~\ref{table1}.
The eighth row gives the number of added $(6S,2P_{1/2})$ states which 
complete the basis of quasi-degenerate basis.
Finally, we end up with multiplicities of $40$, $30$, $12$ and $2$ for 
$F_z = 0, \pm 1, \pm 2, \pm 3$, respectively (ninth row).}
\begin{tabular}{ccccc}
\hline
\hline
\stdrule
                & $F_z=0$   &   $F_z=\pm 1$ &    $F_z=\pm 2$ &   $F_z=\pm 3$ \\
\hline
\hline
\stdrule
$(J=\frac32,F=2)$   &    8       &      8    &       6    &      2     \\
\stdrule
$(J=\frac32,F=1)$   &    8       &      6    &       2    &      0     \\
\hline
\stdrule
$(J=\frac32)$       &    16      &     14    &       8    & 	2   \\
\hline
\stdrule
$(J=\frac12,F=1)$   &     8      &      6    &       2    &      0     \\
\stdrule
$(J=\frac12,F=0)$   &     4      &      2    &       0    &      0     \\
\hline
\stdrule
$(J=1/2)$            &    12      &      8    &       2    &      0     \\
\hline
\hline
\stdrule
$(J=\frac12)+(J=\frac32)$  & 28      &     22    &      10    &      2     \\
\hline
\hline
\stdrule
$(6S,2P_{1/2})$ States     &    12      &      8    &       2    &      0     \\
\hline
\hline
\stdrule
Total \#{} of States   & 40      &     30    &      12    &      2     \\
\hline
\hline
\end{tabular}
\end{center}
\end{minipage}
\end{center}
\end{table*}

%
%
\section{\texorpdfstring{$\bm{6P}$--$\bm{2S}$}{6P--2S} Interaction}
\label{sec4}

%
%
\subsection{Selection of the States}

We are now turning our attention to the 
interaction of excited $6P$ states with 
metastable $2S$ atoms.
If we could restrict the basis of states to 
the $6P_{1/2}$, $6P_{3/2}$, and $2S$ states,
just replacing the $1S$ state from the 
previous calculation with the metastable $2S$
states, then the calculation
would be relatively easy. 
However, as it turns out, there is an additional 
complication: namely, the 
$| (6P_{1/2})_A (2S)_B \rangle$ states are energetically 
quasi-degenerate with respect to $| (6S)_A (2P_{1/2})_B \rangle$
states, and removed from each other only by the 
classic $2S$--$2P_{1/2}$ Lamb shift
(the $n=6$ Lamb shift is much smaller).
It is thus necessary to augment the basis of 
states with the $6S$--$2P_{1/2}$ states.
This turns the interaction into a
$6P_{1/2}$--$6P_{3/2}$--($6S$;$2P_{1/2}$)--$2S$ system,
where the notation indicates that 
the $6S$--$2P_{1/2}$ states are merely added as 
virtual states, for the calculation of second-order
energy shifts. Among the basis states,
the reference states of interest are the 
$6P_{1/2}$ and $6P_{3/2}$ states.

A priori, we have $4 + 8 + 4 + 4 + 4 = 24$ one-atom states
in the $6P_{1/2}$--$6P_{3/2}$--($6S$;$2P_{1/2}$)--$2S$ system,
which amounts to $24^2 = 576$ states for the two atoms,
and the Hamiltonian matrix would have $576^2 = 331776$ entries. 
In order to remain within the quasi-degenerate basis,
we should select only those two-atom states composed 
of an $S$ and a $P$ state,
and only those where the principal quantum numbers add up to 
$2 + 6 = 8$. These selection rules drastically reduce 
the number of states in the basis, 
according to Table~\ref{table3}.
Furthermore, the total Hamiltonian~(\ref{H}) 
commutes with the total angular momentum $\vec F$.
From the $6P_{1/2}$--$6P_{3/2}$--$2S$ system,
we obtain multiplicities of 
$28$, $22$, $10$ and $2$, for the 
manifolds with $F_z = 0$, $F_z = \pm 1$, $F_z = \pm 2$, 
and $F_z = \pm 3$.
By the addition of the $(6S,2P_{1/2})$ states, 
we end up with multiplicities of
$40$, $30$, $12$ and $2$, for the 
manifolds with $F_z = 0$, $F_z = \pm 1$, $F_z = \pm 2$, 
and $F_z = \pm 3$.

Roughly, one proceeds as in the 
$6P_{1/2}$--$6P_{3/2}$--$1S$ system,
sets up the matrices of the total 
Hamiltonian~(\ref{H}) for every $F_z$ manifold,
and then, convinces oneself that the 
averages of the first-order energy shifts
vanish, both for the entirety of states 
within every $F_z$ manifold as well as
for every fine-structure submanifold individually.
Then, one takes as reference states
the $6P$--$2S$ system and evaluates the second-order 
shifts for every state in a given $(J, F, F_z)$ 
manifold. The individual multiplicities 
are given in Table~\ref{table3}.
Observe that even for given $(J, F, F_z)$, the 
$6P_J$--$2S$ reference levels need not all be energetically degenerate
in view of a freedom of choice for the 
hyperfine-resolved complementary $2S$ state in the 
two-atom basis.

\begin{table*}
\begin{center}
\begin{minipage}{0.95\linewidth}
\begin{center}
\caption{\label{table4} Average second-order \vdw{} shifts for $6P_J$ hydrogen atoms
interacting with $2S$ metastable atoms. Entries marked with a long hyphen
(---) indicate unphysical combinations of $F$ and $F_z$ values.
We denote the scaled interatomic distance by $\rho = R/a_0$
and give all energy shifts in atomic units, i.e.,
in units of the Hartree energy $E_h = \alpha^2 m_e c^2$. The notation $\Delta$ is defined in Eq.~(\ref{DefineDelta}).}
\begin{tabular}{c@{\hspace{0.5cm}}c@{\hspace{0.5cm}}c@{\hspace{0.5cm}}c@{\hspace{0.5cm}}c}
\hline
\hline
\stdrule
                & $F_z=0$   &   $F_z= \pm 1$ &    $F_z=\pm 2$ &   $F_z= \pm 3$ \\
\hline
\hline
\lrgrule
$(J=3/2,F=2)$   &  $\dfrac{\num{3.94240e10}}{\rho^6}$
                &  $\dfrac{\num{3.28390e10}}{\rho^6}$
                &  $\dfrac{\num{1.77316e10}}{\rho^6}$ &  0 \\
\lrgrule
$(J=3/2,F=1)$   &  $\Delta+\dfrac{\num{3.49269e10}}{\rho^6}$
                &  $\Delta+\dfrac{\num{2.63584e10}}{\rho^6}$
                &  $\Delta+\dfrac{\num{8.53151e9}}{\rho^6}$ &  --- \\
\lrgrule
$(J=1/2,F=1)$   &  $-\Delta+\dfrac{\num{4.26261e10}}{\rho^6}$
                &  $-\Delta+\dfrac{\num{3.74479e10}}{\rho^6}$
                &  $-\Delta+\dfrac{\num{2.37505e10}}{\rho^6}$ & --- \\
\lrgrule
$(J=1/2,F=0)$   & $\dfrac{\num{3.87153e10}}{\rho^6}$
                & $\dfrac{\num{3.53658e10}}{\rho^6}$
                & ---  &  --- \\
\hline
\hline
\end{tabular}
\end{center}
\end{minipage}
\end{center}
\end{table*}

%
%
\subsection{Second--Order Energy Shifts}
\label{sec4B}

For the states within the individual $(J, F, F_z)$ manifolds,
the averaged second-order energy shifts 
of the $6P_J$--$2S$ reference levels are given in Table~\ref{table4}.
As a function of $J$ and $F$, an averaging over the magnetic quantum
numbers $F_z$ leads to the results
\begin{subequations}
\begin{align}
\left< E(6P_{1/2}, F=0 ) \right>_{F_z} =& \; 
\frac{\num{3.70405e10}}{\rho^6} \, E_h \,, \\
\left< E(6P_{1/2}, F=1 ) \right>_{F_z} =& \; 
\left( -\Delta + \frac{\num{3.68914e10}}{\rho^6} \right) E_h \,, \\
\left< E(6P_{3/2}, F=1 ) \right>_{F_z} =& \; 
\left( \Delta + \frac{\num{2.62434e10}}{\rho^6} \right) E_h \,, \\
\left< E(6P_{3/2}, F=2 ) \right>_{F_z} =& \; 
\frac{\num{2.63399e10}}{\rho^6} \, E_h \,.
\end{align}
\end{subequations}
These results are obtained from the entries in Table~\ref{table4},
using the weighting factors from Table~\ref{table3}
for an averaging over the rows.

One can also average over the possible orientations of $F$,
namely, $F = J \pm \frac12$, for given $J$ and $F_z$.
This amounts to an averaging over the first two entries in the 
columns (for $J = 1/2$), and the third and fourth entry 
(for $J = 3/2$) in every column of Table~\ref{table4}.
The results of the latter procedure are
\begin{subequations}
\begin{align}
\left< E(6P_{1/2}, F_z=0 ) \right>_F = & \;
\left( -\frac23  \Delta + \frac{\num{4.13232e10}}{\rho^6} \right) E_h \,, \\
\left< E(6P_{1/2}, F_z=\pm 1 ) \right>_F = & \;
\left( -\frac34  \Delta + \frac{\num{3.69274e10}}{\rho^6} \right) E_h , \\
\left< E(6P_{1/2}, F_z=\pm 2 ) \right>_F = & \;
\left( -\Delta + \frac{\num{2.37505e10}}{\rho^6} \right) E_h \,, 
\end{align}
\end{subequations}
and
\begin{subequations}
\begin{align}
\left< E(6P_{3/2}, F_z=0 ) \right>_F =& \;
\left( \frac12 \, \Delta + \frac{\num{3.71754e10}}{\rho^6} \right) E_h \,, \\
\left< E(6P_{3/2}, F_z=\pm 1 ) \right>_F =& 
\left( \frac37 \, \Delta + \frac{\num{3.00616e10}}{\rho^6} \right) E_h \,, \\
\left< E(6P_{3/2}, F_z=\pm 2 ) \right>_F =& 
\left( \frac14 \, \Delta + \frac{\num{1.54315e10}}{\rho^6} \right) E_h \,, \\
\left< E(6P_{3/2}, F_z=\pm 3 ) \right>_F =& \; 0 \; \,.
\end{align}
\end{subequations}
As a function of $J$, averaging over both $F$ and $F_z$ leads to the
results
\begin{subequations}
\begin{align}
\label{6P2S_6P12}
\left< E(6P_{1/2} ) \right>_{F, F_z} =& \; 
\left( -\frac34 \, \Delta + \frac{\num{3.69287e10}}{\rho^6} \right) E_h \,, \\
\label{6P2S_6P32}
\left< E(6P_{3/2} ) \right>_{F, F_z} =& \; 
\left( \frac38 \, \Delta + \frac{\num{2.632037e10}}{\rho^6} \right) E_h \,.
\end{align}
\end{subequations}
Without hyperfine resolution, there are four $J = 3/2$ states
and two $J = 1/2$ states.
Hence, an additional average over the fine-structure levels
leads to a cancellation of the term proportional to $\Delta$,
but the $1/\rho^6$ energy shift remains as an overall 
repulsive interaction among $6P$--$2S$ atoms.

We had seen in Eqs.~(\ref{6P1S_6P12}) and 
Eqs.~(\ref{6P1S_6P32}) that the 
fine-structure average of the \vdw{} interaction vanishes 
for the $6P$--$1S$ interaction, in the quasi-degenerate 
basis which we are using.
For the $6P_{1/2}$--$2S$ and $6P_{3/2}$--$2S$ systems,
the \vdw{} interactions are found to be repulsive 
and the \vdw{} coefficients are of order $10^{10}$ in atomic units
[see Eqs.~(\ref{6P2S_6P12}) and~(\ref{6P2S_6P32})].
The latter, comparatively large \vdw{} coefficients mainly are due to the 
virtual $(6S;2P_{1/2})$ states which have to be
added to the quasi-degenerate basis if the 
spectator atom is in the metastable $2S$ state.

%
%
\section{Time Dependence and Oscillatory Terms}

A few remarks on the role of the time dependence of the
interaction and the no-retardation approximation are in order,
in view of recent 
works~\cite{BeDu1997,Be2015,DoGuLa2015,Do2016,JeAdDe2017prl,JeDe2017vdw0},
part of which discuss time-dependent effects in \vdw{} interactions.
First of all, let us emphasize that the Hamiltonian~\eqref{H} 
is manifestly time-independent.
As such, it cannot generate oscillatory terms 
in energy shifts, for reasons of principle.
In order to see this, let us consider the time evolution of a 
matrix element $M = \langle \phi | H | \phi' \rangle$
under the action of $H$, where $| \phi \rangle$ and $| \phi'\rangle$ are 
arbitrary basis states. The matrix element transforms into 
$M(t) = \langle \phi | \exp(\ii H t/\hbar) \, 
H \,  \exp(- \ii H t/\hbar) | \phi' \rangle = M(0) = M$
and therefore is time-independent.
The time-independent Hamiltonian matrix therefore 
has stationary eigenvalues, which describe the 
time-independent energy eigenvalues of the system.
This approach is canonically taken in 
the analysis of the \vdw{} interaction within manifolds
of quasi-degenerate states (see also Ref.~\cite{JoEtAl2002}).

The time-independence of our Hamiltonian~\eqref{HvdW} corresponds to the 
non-retardation approximation for the \vdw{} approximation.
Just like in our recent paper~\cite{JeEtAl2017vdWii},
the validity of the non-retardation approximation,
is tied to the energetic (quasi-)degeneracy of the levels 
in our hyperfine-resolved basis.
In general, retardation sets in when the phase of the virtual
oscillation of an atom changes appreciably over the time that it takes light to
travel from one atom to the other, and back [see the discussion surrounding
Eqs.~(8) and (9) of Ref.~\cite{AdEtAl2017vdWi}].
For our case, the transition wavelengths
correspond to fine-structure, Lamb shift and 
hyperfine-structure transitions. 
The largest of these is the fine-structure interval
$E_{\rm FS} \sim \alpha^4 \, m_e \, c^2$.
The validity of the non-retardation approximation is 
thus equivalent to the condition
\begin{equation} 
\frac{R}{c} \ll  \frac{\hbar}{E_{\rm FS}}  \,,
\qquad
R \ll \frac{\hbar}{\alpha^4 m_e c} =
\frac{a_0}{\alpha^3} \approx 0.136 \, {\rm mm} \,.
\end{equation}
Thus, in the distance range where retardation becomes relevant,
the overall magnitude of the van der Waals
interaction is completely negligible. The interaction, within the manifold of
quasi-degenerate states,  is thus ``instantaneous'' from the point of view of
virtual transitions, and the no-retardation approximation is justified.

In recent works~\cite{JeAdDe2017prl,JeDe2017vdw0},
a somewhat related but different situation is
considered: An atomic interaction
is treated where it is assumed that there are optical transitions available, to
energetically lower states as compared to the excited reference state, which
act as ``virtual resonant transitions'' and lead to long-range oscillatory tails
in the \vdw{} interaction, where the oscillations 
are functions of the interatomic distance (not of time).
This is the case, e.g., for an excited $3S$--$1S$
system, with the virtual $2P$ states in the $3S$ atom acting as virtual resonant
states [see also Ref.~\cite{AdEtAl2017vdWiii}].

Additionally, 
in Refs.~\cite{BeDu1997,Be2015,DoGuLa2015,Do2016},
time-dependent effects have been studied in the context 
of retarded \vdw{} interactions for non-identical atoms.
A comparison to formulas given in Refs.~\cite{DoGuLa2015}
and~\cite{Do2016}, and~\cite{BeDu1997,Be2015},
is in order. 
First, one should observe that the
second-order shifts investigated in
Ref.~\cite{DoGuLa2015} diverge
in the limit of identical atoms, i.e.,
for the case of the ``interatomic
detuning'' $\Delta_{AB} \to 0$.
Also, the energy shifts given in Eqs.~(6) and (7) of
Ref.~\cite{Do2016} diverge, because
(in the notation of Ref.~\cite{Do2016}) one has
$\calU_{ijpq} \propto \mu^A_i \mu^A_j \mu^B_p \mu^B_q/ \Delta_{AB}$
where $\calU_{ijpq}$ is a tensor that enters the calculation of the
energy shift, and the $\mu_i$ are the dipole moment operators
of non-identical atoms $A$ and $B$.
For the case of perfect degeneracy,
we thus have to calculate one-photon rather than
two-photon exchange (in the non-retardation approximation).
This has been done here and in our recent work~\cite{JeEtAl2017vdWii}.
We find, in full analogy to the discussion in Ref.~\cite{JeEtAl2017vdWii},
that the average first-order shifts of the hyperfine-resolved
levels, in first order of the \vdw{} Hamiltonian, vanish.

The interaction with the 
quantized modes of the radiation field 
(which mediate the \vdw{} interaction) is 
``switched on'' at distance $R$ 
and time $t = 0$ in Refs.~\cite{DoGuLa2015,Do2016}.
We here refrain from a discussion on the relevance of this 
approximation. Otherwise, the interatomic interaction energy otherwise 
increases as the atoms approach each other from infinity 
to a finite distance $R$.
We only explore the consequences of the results reported 
in Refs.~\cite{DoGuLa2015,Do2016}.
Let $|\Psi \rangle = |6P\rangle_A | 1S \rangle_B$
denote a specific state within the
$6P$-$1S$ hyperfine manifold, and let
$|\Psi' \rangle = |1S' \rangle_A | 6P' \rangle_B$ denote a different state,
displaced by an energy shift of order %
\begin{equation}
\Delta_{\rm HFS} \sim \Delta_{AB} \sim \calH \,.
\end{equation}
We concentrate on the second-order
shifts due to quasi-degenerate levels
$|\Psi'\rangle = |1S'\rangle_A | 6P' \rangle_B$,
in which case the formalism of~\cite{Do2016}
and~\cite{Be2015} becomes applicable.
In Eqs.~(20) and (6) of Ref.~\cite{Do2016},
it is claimed
that the ``usual'' result for the \vdw{} energy shift
holds only in the limit
\begin{equation}
\Gamma \ll \hbar \Omega \ll \Delta_{AB} \,,
\end{equation}
where $\Gamma$ is the natural line width of the state
(expressed in units of energy),
$\Omega$ is the Rabi frequency of the excitation,
and $\Delta_{AB}$ is the detuning $E_{\Psi'} - E_\Psi$.
This condition cannot be met in our setting
for any Rabi frequency $\Omega$, because we evidently have
$\Gamma \gtrsim \Delta_{\rm HFS} \sim \Delta_{AB} $ (the
natural line width of the $6P$ states
is greater than the hyperfine splitting).
Otherwise, time-dependent oscillatory terms are
obtained in Ref.~\cite{Do2016}.

Likewise, it  is claimed in Ref.~\cite{DoGuLa2015} that
upon a sudden excitation of atom $A$ at $t=0$, at 
a later time $T$,
there are oscillatory terms in the energy which
for our case would be proportional to
\begin{equation}
\cos\left( \frac{2 (E_{6P} - E_{1S}) R}{\hbar c} +
\frac{\Delta_{AB} T}{\hbar} \right) \,,
\end{equation}
and thus oscillatory in both space and time.
These terms are claimed to influence the energy shift dynamically,
after an observation time $T$. Note that the first term on the right-hand
side of Eq.~(4) of Ref.~\cite{DoGuLa2015}
reproduces the usual second-order \vdw{} contribution
(proportional to $R^{-6}$)
to an energy shift in the limit 
of a vanishing wave vector for the transition 
of atom $A$, namely, $k_A \to 0$ (in the notation of Ref.~\cite{DoGuLa2015}),
which is relevant for our quasi-degenerate
manifolds. This term is non-oscillatory in $T$
and gives the leading contribution
to the interaction energy for quasi-degenerate
systems in the non-retardation limit,
as also remarked in Ref.~\cite{JeAdDe2017prl}.

In any case, the additional oscillatory
terms obtained in Eqs.~(20) and (6) of Ref.~\cite{Do2016}
and in Eq. (4) of Ref.~\cite{Do2016}
average out to zero over the observation time $T$.
If we are to evaluate position-dependent
energy shifts (pressure shifts) within an atomic beam,
then we do not know the time $T$ at which an atomic collision
occurs, within the beam, after excitation.
The oscillatory terms in the energy shifts
thus average out to zero, in the calculation
of the pressure shifts due to atomic collisions
within the beam.

In order to avoid ``switching on'' the interaction
with the quantized modes of the radiation field,
suddenly at time $t = 0$, one generally assumes the excited state to be
an asymptotic state (in the context of the $S$-matrix
formalism, see Refs.~\cite{ItZu1980,BeLiPi1982vol4}).
For a didactic presentation of the application 
of this formalism to the ac (``alternating-current'',
oscillatory-field) Stark energy shift
due to an oscillatory external laser field,
see Ref.~\cite{HaJeKe2006}.
Specifically, for manifestly oscillatory terms in the 
interaction Hamiltonian such as a laser field, 
or the quantized electromagnetic field,
one damps the interaction infinitesimally at 
infinity and uses an infinitesimally damped 
time-evolution operator [see Eq.~(21) of Ref.~\cite{HaJeKe2006}]
in order to formulate the energy 
shift within the Gell--Mann--Low theorem.
Or, one matches the $S$-matrix amplitude with the 
energy shift generated by the interatomic 
interaction~\cite{BeLiPi1982vol4,JeAdDe2017prl,JeDe2017vdw0}.
Finally, the calculation of the pressure shift within an 
atomic beam, using the time-independent \vdw{} potentials as input,
is discussed in Chaps.~36 and~37 of Ref.~\cite{So1972}
(within the impact approximation).

%
%
\section{Conclusions}
\label{sec8}

In this paper, we have studied the \vdw{} interaction
of excited $6P$ hydrogen atoms with
ground-state $1S$ and metastable $2S$ atoms.
Within our hyperfine-resolved basis,
in order to obtain reliable estimates of the 
\vdw{} interaction coefficients, one needs 
to consider all off-diagonal matrix elements 
of the \vdw{} interaction Hamiltonian.
Specifically, for hydrogen, the nuclear spin $I = \frac12$ needs to 
be added to the total electron angular momentum $J$, 
resulting in states with the total 
angular momentum $F = J \pm \frac12$ of
electron$+$nucleus.
The explicit construction of the 
hyperfine-resolved states is discussed in Sec.~\ref{sec2}.
For the $6P$--$1S$ system, one needs to 
include both the $6P_{1/2}$ as well as the 
$6P_{3/2}$ states in the quasi-degenerate basis,
because the $6P$ fine-structure frequency is 
commensurate with the $1S$ hyperfine transition 
splitting (see Sec.~\ref{sec3}).
The matrix elements of the total 
Hamiltonian involve the so-called hyperfine--fine--structure 
mixing term (see Sec.~\ref{sec3}),
which couples the $6P_{1/2}(F=1)$ to the 
$6P_{3/2}(F=1)$ levels [see Eq.~(\ref{HFSFSmix})].

The explicit matrices of the total Hamiltonian~(\ref{H})
in the manifolds with $F_z = 3,2$ are 
described in Secs.~\ref{sec3B} and~\ref{sec3C}.
Due to mixing terms of first order in the
\vdw{} interaction
between degenerate states in the two-atom system,
the leading term in the \vdw{} energy,
upon rediagonalization of the Hamiltonian matrix,
is of order $1/R^3$ for the
$6P$--$1S$ interaction, but it averages out 
to zero over the magnetic projections.
The phenomenologically important second-order 
shifts of the energy levels are given in Sec.~\ref{sec3E},
with various averaging procedures illustrating the 
dependence of the shifts on the quantum numbers,
and the dependence of the repulsive or 
attractive character of the interaction on the 
hyperfine-resolved levels.

The same procedure is applied to the 
$6P$--$2S$ interaction in Sec.~\ref{sec4},
with the additional complication that 
virtual quasi-degenerate $(6S;2P_{1/2})$
also need to be included in the basis.
The treatment of the $6P$--$1S$ and 
$6P$--$2S$ long-range interactions reveals the 
presence of numerically large coefficients multiplying the 
$1/\rho^6$ interaction terms, due to the presence 
of quasi-degenerate levels.
The interaction remains nonretarded over all 
phenomenologically relevant distance scales.
The repulsive character of the $6P$--$2S$ interaction
due to the quasi-degenerate virtual $(6S;2P_{1/2})$ levels
is obtained as a surprise conclusion from the 
current investigation.

\section*{Acknowledgments}

This research has been supported by the 
National Science Foundation (Grants PHY--1403973
and PHY--1710856), as well as the Missouri Research Board.

\end{document}